\documentstyle[12pt]{article}

\def\be{\begin{equation}}
\def\ee{\end{equation}}
\def\bea{\begin{eqnarray}}
\def\eea{\end{eqnarray}}

\def\br{\begin{thebibliography}{abc}}
\def\er{\end{thebibliography}}
\def\rf{\bibitem}

\def\a{\alpha}
\def\b{\beta}

\def\d{\delta}
\def\e{\epsilon}

\def\g{\gamma}

\def\l{\lambda}

\def\w{\omega}
\def\p{\pi}

\def\s{\sigma}
\def\t{\tau}

\def\G{\Gamma}

\def\T{{\bf T}}
\def\IC{\relax\,\hbox{$\inbar\kern-.3em{\rm C}$}}
\def\ID{\relax{\rm I\kern-.18em D}}
\def\IF{\relax{\rm I\kern-.18em F}}
\def\IH{\relax{\rm I\kern-.18em H}}
\def\II{\relax{\rm I\kern-.17em I}}
\def\I1{\relax{\rm 1\kern-.28em l}}
\def\IN{\relax{\rm I\kern-.18em N}}
\def\IP{\relax{\rm I\kern-.18em P}}
\def\IQ{\relax\,\hbox{$\inbar\kern-.3em{\rm Q}$}}
\def\IZ{\relax\,\hbox{$\inbar\kern-.3em{\rm Z}$}}
\def\R{\relax{\rm I\kern-.18em R}}
\font\cmss=cmss10 \font\cmsss=cmss10 at 7pt
\def\Z{\relax\ifmmode\mathchoice
{\hbox{\cmss Z\kern-.4em Z}}{\hbox{\cmss Z\kern-.4em Z}}
{\lower.9pt\hbox{\cmsss Z\kern-.4em Z}}
{\lower1.2pt\hbox{\cmsss Z\kern-.4em Z}}\else{\cmss Z\kern-.4em
Z}\fi}

\def\cd{{\cal D}}

\def\rt{\rightarrow}

\def\bar#1{\overline{#1}}

\def\Hat#1{\rlap{\kern.10em$\widehat{\phantom G}$}#1}
\def\HAt#1{\rlap{\kern.05em$\widehat{\phantom G}$}#1}

\def\czp#1{\rlap{\kern.1em$\widehat{\phantom{G\vrule height.8em}}$}#1{}}
\def\Czp#1{\rlap{\kern.05em$\widehat{\phantom{G\vrule height.8em}}$}#1{}}

\newcommand{\sect}[1]{\setcounter{equation}{0}\section{#1}}
\newcommand{\subsect}[1]{\subsection{#1}}

\def\fn{\footnote}
\def\sxn#1{\bigskip\medskip \sect{#1} \smallskip
                                                 }
\def\subsxn#1{\medskip \subsect{#1} \smallskip
                                                }

\begin{document}

\thispagestyle{empty}
\setcounter{page}{0}

\begin{flushright}
UICHEP-TH/95-3 \\
May 1995\\
hep-th/9505028
\end{flushright}

\vspace{1cm}

\begin{center}
{\LARGE SELF-INTERSECTION NUMBERS AND }\\
\vspace{5mm}
{\LARGE  RANDOM SURFACES ON THE LATTICE}\\
\vspace{5mm}
\vspace{0.75cm}
{\large P. Teotonio-Sobrinho
\fn[1]{e-mail: {\tt teotonio@tigger.cc.uic.edu}}}\\
\vspace{.50cm}
{\it Department of Physics, University of Illinois at Chicago,\\
			Chicago, IL 60607-7059, USA.}
\end{center}
\vspace{2mm}

\vspace{.2cm}
\begin{abstract}

String theory in 4 dimensions has the unique
feature that a topological term, the oriented self-intersection number, can be
added to the usual action.
It has been suggested that the corresponding theory of random surfaces
wold be free from the problem encountered in the scaling of the
string tension.
Unfortunately, in the usual dynamical triangulation it is not clear how
to write such a term. We show that for random surfaces on a hypercubic lattice
however, the analogue of the oriented self-intersection number $I[\s]$
can be defined and computed in a straightforward way. Furthermore, $I[\s]$
has a genuine topological meaning in the sense that it
is invariant under the discrete analogue of continuous deformations. The
resulting random surface model is no longer free and may lead to a non trivial
continuum limit.

\end{abstract}

\newpage

\sxn{Introduction}

String theory is supposed to play a crucial role in our understanding of
fundamental physics. Either as an effective model for strong interactions or
as a fundamental theory for unification, it has been the focus of
numerous investigations.
In its Euclidean version, string theory is the statistical mechanics of random
surfaces. A 2d manifold  $M$ is immersed on a target space $X$
resulting on a surface $s\subset X$. The partition function is given by a
weighted integral over all surfaces $s$. Theories of random surfaces are
naturally related to many different physical
systems such as membranes in biophysics \cite{D} and 3d Ising model
\cite{3dI1,3dI2}.
{}From this perspective, there is enough
motivation to investigate lattice versions of random surfaces and
study their continuum limit. From the QCD point of view, it would be very
interesting to have a lattice version of 4-dimensional strings
with a nontrivial continuum limit. Unfortunately, such goal seems to be
very difficult.

In the past few years much work has been done in this direction.
The simplest and most natural model is the immediate
translation of the Nambu-Goto theory. One starts by replacing $X$
by a $\Z^d$ lattice and surfaces by polyhedra made of 2
dimensional plaquettes on $\Z^d$. The partition function is
then defined to be the
sum over all surfaces $\s$ with a statistical weight given by
the number of plaquettes of $\s$. If $M$ is assumed to have the topology
of a sphere, the model is called planar random surfaces \cite{W}-\cite{DFJ2}.
Such  a model  is
directly related to the $1/n$ limit of $SU(n)$ lattice Yang-Mills theory.
Surprisingly, the  planar random surface model
was proved to be trivial \cite{DFJ2} and it can not describe any QCD
physics.

Another approach, called dynamical triangulation \cite{DyTr}, is used to
discretize
Polyakov's string theory. The base manifold $M$ is replaced  by a
generic triangulation where the lengths of the links are taken to be
equal.
The embedding of a given triangulation on a continuous manifold
$X$ defines a surface $s$ and the
action (Gaussian ) can be taken to be the area of $s$. In addition to
the sum over immersions, one also sums over all possible triangulations in
order to take into account the intrinsic geometry of $M$.
The important question is whether the model has a well defined
continuum limit. It has been shown that the string tension does not tend to
zero at the critical point, giving rise to pathologically
crumpled surfaces. Consequently this simple model does not lead to a
sensible continuum limit \cite{AD}.
A natural attempt to overcome the problem is to add to the action a
term depending on the extrinsic curvature of $s$ in order to suppress the
contributions of "spiked" surfaces.
Analytic calculations, using Nambu-Goto action
plus an extrinsic curvature term, suggest that the corresponding
coupling constant renormalizes to zero and consequently the discrete
action can not have a nontrivial continuum limit \cite{H}-\cite{K}.
On the other hand much  numerical work has been
carried out  to simulate Polyakov's action together with extrinsic
curvature terms \cite{ExtCurv}. Some evidence of scaling has been found.

Adding an extrinsic curvature term is not the only way of modifying
the usual Gaussian theory.
When the target space $X$ is 4-dimensional, a new possibility is available for
string theory. It has been shown that a topological term can be added to the
usual string action \cite{BLS} (See also \cite{P,MN}).
It introduces an extra weight factor
given by $exp(i\theta I[\s])$, where $I[s]\in \Z$ is a
topological number. In  a sense, it
is the analogue of the $\theta $-term in QCD. The integer number $I[s]$, the
so-called oriented self-intersection number, is a measure of how the embedding
of $M$ on $X$ self-intersects. The resulting theory is described by the
partition function
\be
   Z(\l,\theta )=\int \cd s~e^{-\l A[s]+i \theta I[s]},\label{2.2}
\ee
where $A[s]$ is the usual Nambu-Goto term given by the area of $s$.
It has been suggested \cite{P} that such a partition
function would describe smooth surfaces for $\theta  =\pi $. Therefore it
would be a better candidate for an
effective theory of QCD. The presence of the analogue of a $\theta $-term
is a very suggestive indication \cite{MN}.

One way of studying (\ref{2.2}) is to introduce a lattice regularization and
make the functional integral into a sum. Unfortunately it is
not so clear how to proceed due to the presence of a topological term.
This is a problem
common to many theories involving  topological terms. The first difficulty
is to define and compute the corresponding counterparts on the lattice.
Secondly, the topological meaning of such terms in a discrete setting is not
always clear. A good example of the situation is given by the QCD instanton
number on the lattice. We refer to \cite{L} for the discussion of one
possible solution to the problem of instanton number.

We would like to have a scheme of discretization for the random surface
problem where the definition of oriented self-intersection number
$I[s]$ naturally corresponds to the continuum counterpart. If the base space
$M$ is
discretized, as it it is in dynamical triangulation, it seems to
be very difficult, or even impossible, to come out with a discrete counterpart
for topological numbers. The reason being that the usual way
of discretizing  the manifold $M$ is
by looking at lattices, i.e. a cell decomposition $K(M)$
made of vertices, links and faces. However, the scalar field describing the
string is defined only for the set $K_0(M)$ of all vertices.
Consequently, the set of
all field configuration is just $\G=X^{n (K_0(M))}$, where $n (K_0(M))$ is
the  number of vertices. The quantity $I[s]$ is a function on $\G $, but
unfortunately $\G $ has no information about $M$. The situation is clearly not
satisfactory and one should try something different.

Some time ago, an alternative approach to discretization was formulated by
Sorkin \cite{S}. In this scheme, $M$ is substituted
by a finite topological space
$Q(M)$ that has the ability of reproducing important topological features of
$M$. When the number of points in $Q(M)$ increases, $Q(M)$
approximates $M$ better and better. It is possible to define a
certain continuum limit, where $M$ can be recovered exactly.
Subsequent research developed this methods and made then usable for doing
approximations in quantum physics \cite{Poset}.
It turns out that this
techniques can be nicely applied to self-intersecting surfaces. Indeed, as it
will be explained,
the corresponding definition of $I[s]$ for the discrete theory
is a faithful translation of the definition  in the continuum.
Furthermore, it  has a truly topological meaning.

We learned from previous work that there
are two alternatives for discretization in terms of finite topological spaces.
In the first approach, both the base space
$M$ and the target space $X$ are replaced by the discrete spaces
$Q(M)$ and $Q(X)$.
In the second one, only the target space $X$ is discretized.
The second possibility is not very useful for a generic field theory,
however it
can be efficiently applied to string theory. In this paper we adopt the second
possibility with additional restrictions on $Q(X)$. Under these specific
conditions the
resulting formalism can be reinterpreted in terms of random surfaces made
of plaquettes embedded on a usual hypercubic lattice. Although this
work was inspired by looking at finite topological spaces, they will not be
explicitly mentioned here. An account of self-intersection numbers when both
target and base spaces are discretized will be reported elsewhere.

In this paper, we present a discrete model for random surfaces corresponding
to the Nambu-Goto theory modified by the presence of the topological term.
The case of surfaces with no handles is a modification of  the usual
planar surfaces. We argue that the pathological behavior
observed for the Gaussian action, i.e., $\theta =0$ in (\ref{2.2}),
may not occur for other values of $\theta $, possibly leading to a nontrivial
continuum limit.

The discretization of (\ref{2.2}), without the term $exp(i\theta I[\s])$,
has been extensively studied in the past \cite{W}-\cite{DFJ2}.
Our main objective in this paper is to include the topological term, or in
other words, to make sense
of the self-intersection number $I[\s]$ for any configuration
in the model. We will
show that $I[\s]$ has all the properties that we want. It is an integer,
gives the right answer in the continuum limit, and it has a topological
meaning in the
sense that it is invariant under the analogue of continuous deformations of
$\s$.

To make the paper self contained, the usual self-intersection number for
the continuum case is reviewed in Section 2. For the same purpose, some
elements of the
theory of cell complexes and homology are briefly mentioned in Section 3.
The discrete model is discussed in Section 4. The self-intersection number
is first defined for a very special class of configurations. Finally,
the extension of $I[\s]$ for an arbitrary configuration is given by an
explicit formula. The topological invariance of $I[\s]$ is also demonstrated.
Some generic comments on the consequences of the term $I[\s]$ are collected in
Section 5.

\sxn{The Usual Intersection Number}\label{se:2}

Consider a 2d manifold $M$ without boundary (parameter space) and  a fixed
4d target manifold $X$. For simplicity one can take $X$ to be
$\R^4$. Let $\varphi :M\rt X$ be a continuous map (immersion) and
$s\subset X$ the surface determined by
$\varphi $. Different points of $M$ can be mapped to the same point of
$X$. Therefore the surface $s$ can have self-intersections. The
self-intersection number $I [s]$ is a measure of how $s$
self-intersects.
Usually $I [s]$ is given in terms of local
fields. Let $\xi _a,\! (a=1,2)$
be local coordinates of $M$ and $\varphi ^\mu , \! (\mu =1,2,3,4)$ the
components of $\varphi $. Then the integer $I[s]$ is given by
\cite{BLS,P,MN}
\be
I [s]=\frac{-1}{16\p }\int d^2\xi \sqrt{g}
g^{ab}\nabla _at^{\mu \nu}
      \nabla _b\tilde t^{\mu \nu} \label{I}
\ee
where
\[
g_{ab}=\frac{\partial \varphi ^\mu }{\partial \xi ^a}
	\frac{\partial \varphi ^\mu }{\partial \xi ^b}\, ,
\]
\[
t^{\mu \nu }=\frac{\epsilon ^{ab}}{\sqrt g}\partial _a\varphi ^\mu
	\partial _b\varphi ^\nu
\]
and
\[
 \tilde t^{\mu \nu }=\frac{1}{2}
	\epsilon ^{\mu \nu \a \b }t^{\a \b}.
\]
If $s$ and $s' $ are homotopic, i.e.
they can be continuously deformed into each other, then $I[s ]=I[s ' ]$.
We will use the notation $s \sim s '$ to indicate homotopy.

The intuitive notion of self-intersection number is very simple. For a
2d surface in 4 dimensions, self-intersection can
happen on regions of dimension
two, one and zero. Suppose $s$ self-intersects only at a certain number
$n$ of isolated points. Furthermore, assume that at any intersection point the
two branches of $s$ are not tangent to each other. In this case
we say that $s$ is transversal. The simplest invariant
associated with $s$ is $I_2[s]$, or intersection module 2. $I_2[s]$ is zero
or 1 if  $n$ is respectively even or odd.  Given any
surface $s '$ with transversal self-intersection, one can show that
$I_2[s ']=I_2[s]$ if $s' \sim s$. The invariant $I_2[s]$ is extended to
non transversal configurations in the following way. Find a transversal
surface
$\tilde s \sim s$ and define $I_2[s]$ to be equal to $I_2[\tilde s]$. This
definition is motivated by a theorem stating that $\tilde s $ always exists
and can be made infinitesimally close to $s$ \cite{GP}.

In this paper we will make use of a distinct, but equivalent \cite{LS},
presentation of $I[s ]$. It turns out that the invariant $I[s]$ in
(\ref{I}) can be seen as a refinement of $I_2[s]$. Instead of
simply counting the number of intersections, one associates
``charges'' $\pm 1$ to each intersection and sums over all charges.
Let $W$ be the set of points $x_i\in X$ such that
$\varphi(p)=\varphi(p')=x_i$, for some pair $p,p'\in M$.
Consider also  a positive oriented
base $\{v_1,v_2\}$ of tangent vectors at $p\in M$ and similarly for
$\{w_1,w_2\}$ at $p'\in M$. The map $\varphi $ will induce two sets
$\{v_1',v_2'\}$ and $\{w_1',w_2'\}$ of vectors tangent to $s $ at $x_i$.
We say that $s$ is transversal iff, for all $x_i\in W$, the sets
\be
 B(x_i):=\{v_1',v_2',w_1',w_2'\}\label{2}
\ee
are linearly independent. Therefore, for each $x_i$, $B(x_i)$ is a base of
tangent vectors and defines an
orientation, called product orientation at $x_i$. One can compare
the product orientation for each $x_i$ with the pre-existent orientation of
$X$ and  assign a ``charge'' $+1$ if the orientations agree, and  $-1$
otherwise.
The oriented self-intersection number $I [s]$ is defined to be
the sum of all such ``charges''.
Observe that there is a potential ambiguity in (\ref{2}), because
we can exchange $\{v_1',v_2'\}$ and $\{w_1',w_2'\}$. Obviously, this is not
the case, since it does not affect the orientation of $B(x_i)$.

The definition of transversality presented so far makes use of tangent vectors,
and this is a problem when dealing with discrete spaces. Fortunately, there
is an alternative
way of defining transversality that is more useful for us.
In some coordinate system, a small neighborhood  of an intersection point
$x_i$ can be identified with
an open set $U_i$ of $\R^4$, where $x_i$ sits at the origin. Let us
call $s_i$ and $s_i'$ the two branches of $s \cap U_i$.
Transversality means that we can find a local coordinate system for $U_{i}$
such that the points of $s$ have coordinates  of the form $(y_1,y_2,0,0)$ for
$s_i$ and $(0,0,y_3,y_4)$ for $s_i'$. In other words, $U_{i}$ can
be identified with the Cartesian product $s_i\times s_i'$.
If we give to $s_i\times s_i'$ the product orientation, then
\be
U_i=I[s_i,s_i']~s_i\times s_i'\,,
\ee
where $I[s_i,s_i']=\pm 1$. The oriented self-intersection number is
defined to be \cite{GP}
\be
I [s]=\sum _iI[s_i,s_i']\label{int}
\ee

Formula (\ref{int}) is valid only for transversal surfaces. The
extension of this definition to an arbitrary configuration depends on the
result mentioned before. Two homotopic transversal configurations
have the same $I[s]$, and
for any non-transversal $s$, there is a transversal $\tilde s$ such that
$\tilde s\sim s$. In the same way as for $I_2[s]$, one can safely
define $I[s]$ to be $I[\tilde s]$.

The main advantage of (\ref{int}) is that it can be generalized to the discrete
situation. However, this approach to self-intersection does not give a way
of computing $I[s]$ for non-transversal configurations.
In this sense, the integral
formula (\ref{I}) is more useful, but unfortunately
very difficult to be translated to the lattice.
For this reason, we will work with the discrete version of (\ref{int}).
Finally, in Section \ref{se:4.3} we will give an explicit formula to
compute the self-intersection number for arbitrary configurations.

\sxn{Hypercubic Lattices}

In this section we briefly review some notions of homology theory that we will
need. We refer to \cite{HW} for a systematic exposition.

Abstractly, an $n$-cell $\a_{(n)}$ is a space (of dimension $n$),
together with subspaces \mbox{$\a^i_{(n-1)}\subset \a_{(n)}$} called faces.
The subsets $\a^i_{(n-1)}$ are themselves $(n-1)$-cells, so we can consider
their corresponding $\a^j_{(n-2)}$ faces. The kind of cells that we will be
interested in are regular, meaning that
any $\a^j_{(n-2)}$ belongs to exactly two $(n-1)$-cells in $\a_{(n)}$.
By definition, an
1-cell have only two 0-cells as faces, and a 0-cell has no faces.
A cell complex $K$ of dimension $n$ is defined to be
an union of $n$-cells and it is
totally characterized by its elements $\a_{(k)}^l$ and their inclusion
relations. Therefore, two abstract complexes $K^1$ and
$K^2$ are regarded as identical if there is an one to one map $f:K^1\rt K^2$
that preserves the inclusion relations. It is customary to indicate by
$K_{(p)}\subset K$ the union of all cells of dimension $p$.
Concretely, an (regular) $n$-cell $\a_{(n)}$ and the corresponding
$\a_{(n-1)}^i\subset \a_{(n)}$  can be realized as $n$-dimensional
polygon in $\R^n$ and respective $(n-1)$-dimensional faces.

A cell decomposition of a
$n$-manifold $Y$ is an abstract complex $K(Y)$ of dimension $n$ such that
its concrete realization is homeomorphic to $Y$. An
important property of $K(Y)$ is that any two $(n-1)$-cells belong to at most
two $n$-cells.

Given two abstract cell complexes $K^1$ and $K^2$, one can define the product
cell complex $K^1\times K^2$. The cells of $K^1\times K^2$ are ordered pairs of
cells
\be
\a_{(n+m)}^{i,j}:=\left(\a^i_{(m)},\a^j_{(n)}\right),
 ~~~\a_{(m)}^i\subset K^1~~~
					\a_{(n)}^j\subset K^2
\ee
together with the inclusion relations
\be
\left(\a^k_{(m-1)},\a^l_{(n-1)}\right)\subset
\left(\a^i_{(m)},\a^j_{(n)}\right)~~~\mbox{iff}~~~
\a^k_{(m-1)}\subset \a^i_{(m)}~\mbox{ and }~\a^l_{(n-1)}\subset \a^j_{(n)}.
\label{5.5}
\ee

In this paper, we  will be restricted to consider  $n$-cells that can
be realized
as cubes of dimension $n$.
Abstractly, a cubic  $n$-cell $L_{(n)}$ is by definition the product
\be
L_{(4)}=L^1\times L^2\times ...\times L^n
\ee
of $n$ 1-cells $L^i$. In other words, a cell $\a \subset L_{(n)}$ is given by
\be
\a=(\a^1,\a^2,...,\a^n)
\ee
where $\a^i$ can be $L^i$ or one of its vertices. A cubic cell complex of
dimension $n$ will be the union of cubic cells of dimension $n$.

Given a cell complex $K$, one defines the vector space $C_n(K,\Z)$ as the
linear combination of $n$-cells, with coefficients in $\Z$
\be
C_n(K,\Z)=\left\{ \xi _{(n)} =\sum _i \l_i\a_{(n)}^i~:~~\l_i\in \Z,~~~
     \a^i_{(n)}\subset K\right\}
\ee
The vectors $\xi _{(n)}$ are called $n$-chains. The direct sum of all
$C_n(K,\Z)$ will be denoted by $C(K,\Z)$.

The definition of orientation is related to a linear operator
$$\partial :C_n(K,\Z)\rt C_{(n-1)}(K,\Z),$$
called the boundary operator. It is
enough to define $\partial $ for the base elements $\a^i_{(n)}$.
Intuitively, the boundary $\partial \a^i_{(n)}$ of an
 $n$-cell $\a^i_{(n)}$ has
to do with its faces. In other words, it is  a
linear combination, with coefficients $\pm 1$, of all $(n-1)$-cells
$\a^j_{(n-1)}$ such that $\a^j_{(n-1)}\subset \a^i_{(n)}$. We define
\be
\partial \a^i_{(n)}=0~~ \mbox{ if }n=0
\ee
and
\be
   \partial \a^i_{(n)}=\sum _j I_{nc}(\a^i_{(n)},\a^j_{(n-1)})
\a^j_{(n-1)}.\label{5.2}
\ee
The coefficients $I_{nc}(\a^i_{(n)},\a^j_{(n-1)})=\pm 1$ are called the
incidence numbers and  they have to be assigned in such way that
\be
\partial \partial \xi=0~~~~~\mbox{for any }\xi \in C(K,\Z). \label{5.2.1}
\ee
In terms of incidence numbers, (\ref{5.2.1}) is equivalent to
\be
\sum _j I_{nc}(\a^i_{(n)},\a^j_{(n-1)})I_{nc}
(\a^j_{(n-1)},\a^k_{(n-2)})=0.\label{5.2.2}
\ee
It turns out that incidence numbers can be assigned
recursively in a simple way, and this is what is used to define orientation.
First, it is assumed that the boundary of an 1-cell $\a_{(1)}$ with faces
$\a^1_{(0)}$ and $\a^2_{(0)}$ can only be $\pm (\a^2_{(0)}-\a^1_{(0)})$.
In other
words, for a given $i$ there are only two possibilities for
$I_{nc}(\a_{(1)}^i,\a_{(0)}^j)$, and one is the negative of the other.
Suppose now that
all $I_{nc}(\a_{(1)}^i,\a_{(0)}^j)$ have been chosen for a given 2-cell
 $\a_{(2)}^k$.
It is easy to see that  there are only two  possibilities for
$I_{nc}(\a_{(2)}^k,\a_{(1)}^i)$ satisfying (\ref{5.2.2})
and one is the negative of the
other. This is actually a general fact. Once the incidence numbers are chosen
for the faces $\a_{(n-1)}^i$ of an $n$-cell $\a_{(n)}^k$, there are only
two possible choices for $I_{nc}(\a_{(n)}^k,\a_{(n-1)}^i)$.

{}From above it follows that, once we find a possible configuration of
incidence numbers, all
the others can be obtained by a certain set of transformations. Let us
introduce a function $g(\a)$ from $K(Y)$ to $\{-1,+1\}$. Given a possible
configuration of incidence numbers $I_{nc}^0$ we define a new configuration
$gI_{nc}^0$
\be
gI_{nc}^0(\a^i_{(r)},\a^j_{(r-1)})=g(\a^i_{(r)})~I_{nc}^0(\a^i_{(r)},
\a^j_{(r-1)})~g(\a^j_{(r-1)}).
\label{t}
\ee
It is clear that the $gI_{nc}^0$ satisfy (\ref{5.2.2}). Furthermore, all
possibilities for $I_{nc}$ can be generated in this way, by starting
from any $I_{nc}^0$.

Now consider an $n$-dimensional complex $K(Y)$ associated with some
$n$-manifold $Y$.
What we call local orientations of $K$ is the freedom to choose
independently $I_{nc}(\a^i_{(n)},\a^j_{(n-1)})$ at each $n$-cell.
We say that two configurations $I_{nc}$ and $I_{nc}'$ define the same local
orientation of $K(Y)$, or are equivalent $I_{nc}\sim I_{nc}'$, if
they are related by a transformation (\ref{t}) with $g(\a_{(n)}^i)=1$
\[
I_{nc}\sim I_{nc}'~~\mbox{ iff }~~~I_{nc}'=gI_{nc}~~
\mbox{ for some $g$ such that }g(\a^i_{(n)})=1.
\]
A global orientation for $K(Y)$ appears when we start to compare the local
orientations for neighboring $n$-cells. We say that the local orientation at
$\a^1_{(n)}$ agrees with the local orientation at $\a^2_{(n)}$ iff
\be
   I_{nc}(\a^1_{(n)},f_{(n-1)})=-I_{nc}(\a^2_{(n)},f_{(n-1)}),\label{5.4}
\ee
where $f_{(n-1)}$ is the unique common face. An $n$-dimensional complex,
together with an orientation, is called oriented iff all the local
orientations agree.

A vector $\xi _{(n)}\in C_{(n)}(K,\Z)$ of the form
\be
\xi _{(n)}=\sum_i s_i\a_{(n)}^i~;~~~~s_i=\pm 1\label{vec}
\ee
is interpreted as the oriented $n$-dimensional subcomplex of $K$ given by
\[
\bigcup _i s_i\a_{(n)}^i,
\]
where the factors $s_i=\pm 1$ indicate orientation.
If $\xi _{(n)}$ is globally oriented, then
$\partial \xi _{(n)}$ is also globally oriented.

Let $K^1$ and $K^2$ be two globally oriented cell complexes and
$\partial _1$ and $\partial _2$ be boundary operators defined on $C(K^1,\Z)$
and $C(K^2,\Z)$. Consider an operator $\partial $
acting on \mbox{$C(K^1\times K^2,\Z)$} in the following way
\be
   \partial (\a^1_{(m)},\a^2_{(n)})=(\partial _1\a^1_{(m)},\a^2_{(n)})+
(-1)^m
(\a^1_{(m)},\partial _2\a^2_{(n)})\,,\label{5.6}
\ee
It follows immediately that $\partial ^2=0$. Therefore $I_{nc}^1$ and
$I_{nc}^2$ will
define a configuration of incidence numbers $I_{nc}^1\times I_{nc}^2$ for
$K^1\times K^2$. If $g_1I_{nc}^1$ and $g_2I_{nc}^2$ are two other
incidence numbers equivalent to $I_{nc}^1$ and $I_{nc}^2$, a simple
calculation shows that
\be
g_1I_{nc}^1\times g_2I_{nc}^2=g_{(1\times 2)}I_{nc}^1\times I_{nc}^2
\ee
where $g_{(1\times 2)}\left((\a^1,\a^2) \right)=g_1(\a^1)g_2(\a^2)$. Therefore
(\ref{5.6}) induces a canonical  orientation on $K^1\times K^2$, called the
product orientation. It is a simple exercise to verify that the product
orientation is also global.

An 1-cell, or link $L^i$, is totally determined by its vertices $a^i$
and $b^i$. It is standard to write $[a^i,b^i]$ for $L^i$ and define
\be
\partial [a^i,b^i]= [a^i]-[b^i]\label{315}
\ee
then $[b^i,a^i]$ will be identified with $-L^i$. Therefore, the cube
\be
L_{(4)}=([a^1,b^1],[a^2,b^2],[a^3,b^3],[a^4,b^4])\label{L4}
\ee
has a standard set of incidence number determined by (\ref{315}) and
(\ref{5.6}). Whenever we write a cubic cell as in (\ref{L4}),
the standard incidence numbers are assumed.

\sxn{The Discrete Model}

The discretization is done by introducing a grid on the space $\G$
of all surfaces. This allow us
to write the functional integral (\ref{2.2}) as a sum. In other words,
$\G$ will be substituted by some discrete space $\G_d$, where we can
define an area $A[\s]$ and and an intersection number $I[\s]$ for any
configuration $\s\in \G_d$.

\subsxn{Space of Configurations}\label{s:2}

The space of configurations we need to consider is given by the set
of all immersions $\varphi $ of a 2-dimensional manifold $M$
on some 4-dimensional target space $X$.\fn{We assume that $X$ has no boundary,
but $M$ may have boundary components.}
However, the action
(Nambu-Goto) depends only on the area of the surface $s$ determined by
$\varphi $. Any two immersions that give the
same surface $s$ in $X$ are regarded as equivalent.
The relevant set of configurations $\G$ is then the set of all such
surfaces. Evidently not all
$s$ are submanifolds of $X$. They can be degenerated surfaces
in the sense that they can
fold on themselves, i.e., more than one point of $M$ can be mapped
to the same point of $X$.

Let us assume, for simplicity, that $X=\R^4$.
Consider the discrete lattice $\Z^4$ of points (vertices)
$v_i\in X$ that have integer coordinates in some lattice spacing unit $a$.
It determines a cell decomposition $K(X)$ of $X$ where $K_{(0)}=\Z^4$ and
$K_{(n)}$ is the set of all $n$-dimensional elementary cubes determined by
$\Z^4$.

It is useful to think of $K(X)$ as the product of 1-dimensional
complexes. According to the notation introduced at the end of
Section 3, an arbitrary cell $\a\subset K(X)$ will be written as
\be
\a=(\a^1,\a^2,\a^3,\a^4)\label{6.1}
\ee
where the variable $\a^i$ can take the values $[p^i]$ (0-cell) or
$[p^j,p^j+1]$ (1-cell), $p_i\in \Z$.
They will be the base elements for $C(K(X),\Z)$.
Similarly one can also discretize the 4 dimensional torus $\T^4$ by
taking $K(X)^{(0)}=(\Z_n)^4$. In this paper we will be limited to examine
only these two cases.

We are now in a position to define the discrete space $\G_d$ that will be used
to approximate the infinite dimensional space of configurations $\G$.
The set
$\G_d$ will be a countable sub set of $\G$. A configuration  $\s$ belongs
to $\G_d$ if the corresponding surface lies entirely on plaquettes of $K(X)$.
Since the number of plaquettes is countable, so is the number of elements in
$\G_d$.

Another way of interpreting $\G_d$ is to think of a configuration $\s$ as
the natural two dimensional generalization of a random walk. Let us explain.
Consider a random walker that
starts at a vertex $v^0$ and then moves to a neighboring vertex
$v^1$. The trajectory, or curve traversed by the random walker,  can be
specified by the oriented link $[v^0,v^1]$. The subsequent steps  can be
described by adding more links to one end of the curve.
Eventually, the random walker may
go to a vertex $v$ that has been visited before, and the curve self-intersects.
In this case, the curve is no longer regular, in the sense that $v$ belongs
to more than 2 links.
Analogously, one starts to construct a surface by marking some 2d subcomplex
$K^p\subset K(X)$, where $K^p$ is the union of $p$ plaquettes of $K(X)$.
The surface $K^p$ is supposed to be regular in the sense that all 1-cells
belongs to at most 2 plaquettes. Alternatively, $K^p$ can be also written as
a vector in $C_{(2)}(K(X),\Z)$
\be
K^p=s_1\a_{(2)}^1+s_2\a_{(2)}^2+...+s_p\a_{(2)}^p, ~~~s_i=\pm 1, \label{6.2}
\ee
where $\a_{(2)}^i$ are base elements of the form (\ref{6.1}).
Now,  one tries to add one more plaquette
$s_{p+1}\a_{(2)}^{p+1}$ in such way that it has at least one common
link with $K^p$. Eventually, it may happen that, for the resulting complex,
some links belong now to more than 2 plaquettes and the resulting complex
$K^{p+1}=K^p+s_{p+1}\a_{(2)}^{p+1}$
would not be regular. The extreme case is when
\be
\a_{(2)}^{p+1}=\a_{(2)}^{j}, ~~~\mbox{for some $\a_{(2)}^{j}$ in $K^p$},
\ee
meaning that $\a_{(2)}^{p+1}$ has been previously marked. The idea is to
make $K^{p+1}$ regular by hand.
One enlarges $K(X)$ by introducing a copy of the base element
$\a_{(2)}^{p+1}$ and denoting it by $\bar{\a_{(2)}^{p+1}}$.
Then, $K^{p+1}$ is defined to be the abstract complex given by
\be
K^{p+1}=K^p+s_{p+1}\bar{\a_{(2)}^{p+1}}. \label{6.3}
\ee
and it is  regular by construction. The process is iterated a number of
times. Eventually, it will be necessary to introduce many copies of
a given element $\a_{(2)}^{i}$. They will be denote by $\bar{\a_{(2)}^i}$,
$\bar{\bar {\a_{(2)}^i}}$, etc.

A configuration $\s$ with area $n$ is any abstract cell complex $\s= K^n$
constructed as above such that $\s$ represents a cell decomposition of $M$.
We can write $\s$ in the form
\be
\s=\sum _{i=1}^l\b^i, \label{s}
\ee
where
\[
\b^i=
s_i\a_{(2)}^i+\bar{s_i}\,\bar{\a_{(2)}^i}+
   \bar{\bar{s_i}} \,\, \bar{\bar {\a_{(2)}^i}}+...~.
\]

Notice that a configuration $\s $ can not in general be interpreted as a
subcomplex of $K(X)$, i.e., a vector of the form (\ref{vec}). This happens
only if $\s$ self-intersects on a subcomplex of dimension zero.

There is a very useful map $\xi $ from $\G_d $ to
$C_{(2)}(K(X),\Z)$.
If $\s$ is as in (\ref{s}), then $\xi (\s)$ is defined to be the following
vector in $C_{(2)}(K(X),\Z)$
\be
\xi (\s)=\left(s_1+\bar{s_1}+\bar{\bar {s_1}}+...\right)\a_{(2)}^1 +
\left(s_2+\bar{s_2}+\bar{\bar {s_2}}+...\right)\a_{(2)}^2 +...+
\left(s_l+\bar{s_l}+\bar{\bar {s_l}}+...\right)\a_{(2)}^l. \label{xis}
\ee
In other words, all occurrences of $\bar{\a_{(2)}^i}$,
$\bar{\bar{\a_{(2)}^i}}$, etc, in (\ref{s}) are replaced by $\a_{(2)}^i$.
This map will be used on Section 4.3.

Let $\g_i, (i=1,...,n)$ denote fixed loops on $K(X)$. The relevant
observables are the $n$ point Green functions
\be
Z_{n,m}(\g_1,...,\g_n;\l, \theta )=\sum _{\s\in \G_d(\g_1,...,\g_n)}
e^{-\l A[\s]+i\theta I[\s]}\label{f.1}
\ee
The sum is done over the set $\G_d(\g_1,...,\g_n)$ of all surfaces with
$n$ holes and $m$ handles,  such that
\be
\partial \s=\bigcup _{i=1}^n\g_i.
\ee
Some correlation functions play a special rule in the analysis of the theory.
For example the string tension, is defined by
\be
\t(\l,\theta )=\lim _{\g\rt \infty}\frac{1}{LM}\log {Z_{1,m}
(\g_{LM};\l,\theta )},
\ee
where $\g_{LM}$ is a rectangular loop with $L\times M$ links.

The simplest random surface model
would be given by the sum over surfaces with fixed topology. Let us assume,
for example, $X=\R^4$ and surfaces with no handles. This gives us a
generalization of the planar random surface model.

\subsxn{Intersection Number}\label{se:4.2}

As defined on Section \ref{se:2}, the self-intersection number involves
the notion of transversality. We would like to have a definition of
transversality for our discrete surfaces that is a natural generalization
of the definition for continuous surfaces. Let us consider on $\R^{(n+m)}$ two
hyper surfaces $s^1_{(m)}$ and $s^2_{(n)}$ of dimensions
$m$ and $n$. Suppose they meet at a point $x$.  We say that they are
perpendicular if their tangent vectors are perpendicular. It is also
equivalent to say that for a small neighborhood $U_x$ of $x$, the surfaces
are flat and $U_x$ can be canonically identified with $s^1_{(m)}\times
s^2_{(m)}$. If this is the case, $s^1_{(m)}$ and $s^2_{(n)}$ are surely
transversal. If we are dealing with cubic cells this seems to be the
natural notion of transversality. Let us make the idea more precise.

We will use the convention that $\a^i$, $v^i$ and $l^i$ are variables
related to the 1-dimensional cell $[p^i,p^i+1]$, $p^i\in \Z$, with the
following ranges
\begin{eqnarray} \a^i & = &
[p^i],[p^i],[p^i,p^i+1];\nonumber \\
 v^i & = & [p^i],[p^i+1]; \nonumber \\
 l^i & = & [p^i,p^i+1]. \nonumber
\end{eqnarray}

Consider a cubic cell $\a_{(4)}=(l^1,l^2,l^3,l^4)$ as defined on
Section 3. Let  $V,L$ denote subcells of $\a_{(4)}$. Consider $V$ to be
a 3d cube and $L$ a link  such that they have one common
vertex. For example, one can take
\begin{eqnarray}
V & = & (l^1,l^2,l^3,v^4)\nonumber \\
L & = & (v^1,v^2,v^3,l^4) \label{7.1}
\end{eqnarray}
sharing the vertex $(v^1,v^2,v^3,v^4)$. From
(\ref{7.1}) we see that $V$ and $L$ are perpendicular in the obvious sense.
Arbitrary cells $\w_V$ and $\w_L$ of $V$ and $L$ are of the form
\be
\w_V=(\a^1,\a^2,\a^3,v^4)~~\mbox{ and }~~\w_L=(v^1,v^2,v^3,\a^4).
\ee

There is a canonical one to one map $f$ between the product complex
$V\times L$ and $\a_{(4)}$ given by
\be
f\left((\w_V,\w_L)\right)=(\a^1,\a^2,\a^3,\a^4)\label{5.8}.
\ee
One can show that $f$ preserves the inclusion relations and that
if $V\times L$ is oriented according to (\ref{5.6}), it also preserves
orientation. Consequently $V\times L$ can be identified with $\a_{(4)}$.
We will write
\be
V\times L=\a_{(4)},\label{5.9}
\ee
instead of $f(V\times L)=\a_{(4)}$
since the identification (\ref{5.8}) is canonical. The analogous identification
of $L\times V$ with $\a_{(4)}$ does not preserve orientation, and we write
\be
L\times V=-\a_{(4)}
\ee

The last two formulas can be generalized in an obvious way. First let us
introduce some notation. Consider a cube $\a_{(n+m)}=(l^1,...,l^{n+m})$
of dimension $(n+m)$.
We will write a $n$-dimensional sub-cell belonging to $\a_{(n+m)}$ as
\be
\a_{[a_1a_2...a_n]}= [(l^{a_1},l^{a_2},...,l^{a_n},
	v^{a_{n+1}},...,v^{a_{n+m}})],
\ee
where indices $a_i$ take values $(1,...,n+m)$. The square brackets $[~.~]$
stands for the ordering of the indices $a_1,a_2,...,a_n$, or the ordered
permutation of the objects with indices $a_1,a_2,...,a_n$. The notation
indicates that $\a_{[a_1a_2...a_n]}$ have link components only on the
directions $a_1,...,a_n$.  In other words, a generic cell in
$\a_{[a_1a_2...a_n]}$ is of the form
\be
\w_{[a_1...a_n]}=
           \left[(\a^{a_1},...,\a^{a_n},
	v^{a_{n+1}},...,v^{a_{n+m}})\right].
\ee
Suppose now that we have another subcell $\a_{[b_1...b_m]}$
that shares exactly one vertex with $\a_{[a_1...a_n]}$. Consequently the
two sets of indices $\{a_1...a_n\}$ and $\{b_1...b_m\}$ cannot
have any elements in common. There is a canonical one to one map
\mbox{$f:\a_{[a_1...a_n]}\times \a_{[b_1...b_m]}\rt \a_{(m+n)}$} given by
\be
f((\w_{[a_1...a_n]},\w_{[b_1...b_m]}))=\left[(\a^{a_1},...,\a^{a_n},
\a^{b_1},...,\a^{b_m})\right].
\ee
It is a simple matter to show that under this canonical identification we
can write
\be
\a_{[a_1...a_n]}\times\a_{[b_1...b_m]}=
\epsilon _{[a_1...a_n][b_1...b_m]}\a_{(n+m)}\label{x}
\ee
where $\epsilon _{i...k}$ is the usual Levi-Civita symbol.

Observe that, for two arbitrary cells $\a_{(n)}$ and $\b_{(m)}$, the
product $\a_{n}\times \b_{m}$ always makes sense as an abstract complex.
However, its canonical identification with an $(n+m)$-cell $\a_{(n+m)}$
only makes sense if they belong to $\a_{(n+m)}$ and share a single
vertex.

Let $K(X)$ be the cubic cell decomposition of $\R^4$ (or $\T^4$) and
$\s $ a configuration in $\G_d$. The first condition for $\s$ to be
considered transversal is that it self-intersects only on vertices. In
particular, $\s $ has to be a 2-dimensional subcomplex of $K(X)$.  Let
$v=([p^1],[p^2],[p^3],[p^4])$ be one of the vertices where the self
intersection occurs.  We define a neighborhood $U_v$ of $v$ to be the
union of all 4-cells $\a_{(4)}^k$ that
contain $v$. An example of of such a cell is
$([p^1-1,p^1],[p^2-1,p^2],[p^3,p^3+1],[p^4,p^4+1])$.
Since we are restricted to cell
decompositions of  $\R^4$ (or $\T^4$), $U_v$ is the union of
16 4-cells. In other words, considered as a vector in
$C_{(4)}(K(X),\Z)$, $U_v$ is given by
\be
 U_v=\sum _{k=1}^{16} \a_{(4)}^k,
\ee

Consider the 1-dimensional complex $L^i=[p^i-1,p^i]\cup [p^i,p^i+1]$, or
\be
L^i=[p^i-1,p^i] + [p^i,p^i+1]
\ee
made of 2 adjacent vertices. One can see that neighborhood $U_v$
is the product of 4 such 1-dimensional complexes. In other words,
\be
U_v=(L^1,L^2,L^3,L^4).
\ee

Let $\s_v$ and $\s_v'$ be the two components of $\s\cap U_v$. We say that
the intersection is transversal iff $\s$ and $\s'$ are of the form
\be
\begin{array}{ccc}
\s_v&=&s\left[(L^a,L^b,[p^c],[p^d])\right]\\
    \\
\s_v'&=&s'\left[(L^c,L^d,[p^a],[p^b])\right]
\end{array}~~~\mbox{with $[a,b,c,d]=1234~$ and $~s,s'=\pm 1$}.\label{7.2}
\ee
In this case we can write
\be
U_v=I[\s_v,\s_v']~\s_v\times \s_v'\,,\label{5.11}
\ee
where the canonical identification is being used. The coefficient
$I[\s_v,\s_v']=\pm 1$ is called the intersection number at $v$. From
(\ref{x}), (\ref{7.2}) and (\ref{5.11}), it follows that
\be
I[\s_v,\s_v']=ss'\e _{[ab][cd]}.\label{ss}
\ee

Finally,
the self-intersection number $I[\s]$ is defined to be the sum of all
intersection numbers
\be
I[\s]=\sum_v I[\s_v,\s'_v].\label{5.13}
\ee

Let $s$ be the continuous surface associated with a transversal $\s$.
The surface $s$ is transverse in the usual sense, therefore
$I[s]$ defined by (\ref{int}) can also be computed. It is a very simple
exercise involving tangent vectors to show that $I[s]=I[\s]$.

\subsxn{Topological Invariance and Non-transversal Configurations}
\label{se:4.3}

The definition of the self-intersection number for transversal
configurations, presented in Section \ref{se:4.2}, is the exact analogue
of the continuous definition on Section \ref{se:2}.  To complete the
correspondence with the continuous case we need to introduce on $\G_d$ a
notion of continuous deformations, or homotopy of configurations, and show
that $I[\s]$ is an invariant. We also have to extend the $I[\s]$ to
non-transversal configurations.

Intuitively, two configurations $\s_1$ and $\s_2$ should be considered
homotopic iff their continuous counterparts $s_1$ and $s_2$ can be
deformed into each other by a sequence of small deformations.
Let us explain what is meant by a small deformation. Given
$\s_1\in \G_d$, consider a small portion $D_1$ of $\s_1$ such that $D_1$
is a $2$-dimensional sub-complex in $K(X)$. Furthermore, $D_1$ is required
to be topologically equivalent to a disk. A small deformation will be a
process where $D_1$ is removed and substituted by another disk $D_2\subset
K(X)$. We say that the resulting surface $\s_2$ is a
continuous deformation of $\s_1$
iff $(D_2\cup-D_1)$ is the boundary of a 3d complex $B\subset K(X)$, or
\be
D_2-D_1=\partial B,
\ee
where $B$ is topologically equivalent to a $3$-dimensional ball. It is
clear that a minimal deformation happens when $B$ is a cube, $D_1$ is one
of its plaquettes and $D_2$ the union of the other 5 plaquettes.
However, such a minimal deformation is not enough
to generate all small deformations, as we will illustrate by an example. Let
$D_1$ be the union of two adjacent plaquettes as in Fig. 1(a).
Let us apply to each plaquette of $D_1$ the minimal deformation described
above. The resulting surface (Fig. 1(b)) consists of two cubic boxes,
open on the top, placed side by side. It does not correspond to a regular
surface. Notice that it has 2 superposed plaquettes that are glued along the
top link (see figure). Another alternative is to deform $D_1$ into $D_2'$
(Fig. 1(c)), a regular surface consisting of single box open on the top.
But for consistency, $D_2$ has to be homotopic to
$D_2'$. It is clear that, to have a complete set of minimal deformations,
we need to include another deformation rule. Two superposed plaquettes
glued along some of their links can be removed. Obviously,
the links they do not share should remain. An example of the second type of
elementary deformations is shown in Fig. 1(d).
\begin{figure}[t]
\begin{center}
\unitlength=1.00mm
\linethickness{0.8pt}
\begin{picture}(147.00,120.00)
\put(0.00,20.00){\line(0,1){25.00}}
\put(0.00,45.00){\line(1,0){50.00}}
\put(50.00,45.00){\line(0,-1){25.00}}
\put(50.00,20.00){\line(-1,0){50.00}}
\put(0.00,20.00){\line(1,0){25.00}}
\put(25.00,20.00){\line(0,1){25.00}}
\put(0.00,20.00){\line(1,0){25.00}}
\put(0.00,45.00){\line(2,1){20.00}}
\put(20.00,55.00){\line(1,0){50.00}}
\put(50.00,45.00){\line(2,1){20.00}}
\put(70.00,55.00){\line(0,-1){25.00}}
\put(70.00,30.00){\line(0,0){0.00}}
\put(70.00,30.00){\line(-2,-1){20.00}}
\put(0.00,110.00){\line(1,0){50.00}}
\put(50.00,110.00){\line(2,1){20.00}}
\put(70.00,120.00){\line(-1,0){50.00}}
\put(20.00,120.00){\line(-2,-1){20.00}}
\put(0.00,110.00){\line(0,0){0.00}}
\put(25.00,110.00){\line(2,1){20.00}}
\put(127.00,110.00){\line(2,1){20.00}}
\put(95.00,120.00){\line(-2,-1){20.00}}
\put(75.00,110.00){\line(0,0){0.00}}
\put(100.00,110.00){\line(2,1){20.00}}
\put(75.00,110.00){\line(0,-1){25.00}}
\put(75.00,85.00){\line(1,0){20.00}}
\put(95.00,85.00){\line(1,5){5.00}}
\put(100.00,110.00){\line(1,-5){5.00}}
\put(127.00,85.00){\line(0,1){25.00}}
\put(147.00,120.00){\line(0,-1){25.00}}
\put(147.00,95.00){\line(-2,-1){20.00}}
\put(95.00,120.00){\line(0,-1){8.00}}
\put(80.00,45.00){\line(2,1){20.00}}
\put(75.00,20.00){\line(1,5){5.00}}
\put(80.00,45.00){\line(1,-5){5.00}}
\put(75.00,20.00){\line(5,2){9.00}}
\put(85.00,20.00){\line(2,1){20.00}}
\put(105.00,30.00){\line(-1,5){5.00}}
\put(120.00,20.00){\line(0,1){25.00}}
\put(140.00,55.00){\line(0,-1){25.00}}
\put(120.00,20.00){\line(2,1){20.00}}
\put(120.00,45.00){\circle*{2.00}}
\put(120.00,20.00){\circle*{2.00}}
\put(140.00,30.00){\circle*{2.00}}
\put(140.00,55.00){\circle*{2.00}}
\put(35.00,74.00){\makebox(0,0)[cc]{(a)}}
\put(110.00,74.00){\makebox(0,0)[cc]{(b)}}
\put(110.00,5.00){\makebox(0,0)[cc]{(d)}}
\put(35.00,5.00){\makebox(0,0)[cc]{(c)}}
\put(20.00,55.00){\line(0,-1){9.00}}
\put(45.00,46.00){\line(0,1){9.00}}
\put(95.00,85.00){\line(2,1){8.00}}
\put(75.00,110.00){\line(1,0){51.94}}
\put(95.00,120.00){\line(1,0){52.00}}
\put(105.00,85.00){\line(1,0){21.94}}
\put(120.00,120.00){\line(1,-4){1.90}}
\put(108.00,36.00){\line(1,0){6.99}}
\put(115.00,38.00){\line(-1,0){7.04}}
\put(117.00,37.00){\line(-2,1){4.02}}
\put(113.00,35.00){\line(2,1){4.00}}
\put(5.00,105.00){\makebox(0,0)[cc]{{\large $D_1$}}}
\put(80.00,80.00){\makebox(0,0)[cc]{{\large $D_2$}}}
\put(5.00,15.00){\makebox(0,0)[cc]{{\large $D_2'$}}}
\linethickness{0.4pt}
\put(70.00,30.00){\line(0,0){0.00}}
\put(125.00,95.00){\line(1,0){21.94}}
\put(125.00,95.00){\line(-2,-1){20.00}}
\put(125.00,95.00){\line(-1,6){2.00}}
\put(75.00,85.00){\line(2,1){20.00}}
\put(95.00,95.00){\line(1,0){20.00}}
\put(95.00,95.00){\line(0,1){0.00}}
\put(115.00,95.00){\line(-2,-1){9.97}}
\put(115.00,95.00){\line(1,6){2.03}}
\put(70.00,30.00){\line(-1,0){18.93}}
\put(49.00,30.00){\line(-1,0){23.00}}
\put(24.00,30.00){\line(-1,0){4.04}}
\put(20.00,30.00){\line(-2,-1){20.00}}
\put(20.00,30.00){\line(0,1){0.04}}
\put(45.00,43.00){\line(0,-1){0.03}}
\put(45.00,30.00){\line(-2,-1){19.93}}
\put(95.00,95.00){\line(0,1){14.00}}
\put(45.00,30.00){\line(0,1){14.00}}
\put(20.00,30.00){\line(0,1){14.00}}
\put(120.00,120.00){\line(-1,-5){1.99}}
\end{picture}
\end{center}
{\footnotesize {\bf Fig. 1.} (a) is the disk $D_1$ made of two adjacent
plaquettes. (b) is a possible continuous transformation $D_2$, where each
plaquette of $D_1$ is minimally deformed. It consists of two cubic boxes,
open on the top, placed side by side. The two superposed plaquettes are
drawn slightly separated to make the picture clear. (c) is an alternative
deformation $D_2'$ of $D_1$. The elementary deformation connecting $D_2$
and $D_2'$ is shown in (d) }
\end{figure}

\clearpage

We recall that associated to each configuration $\s\in \G_d$ there is a 2-chain
\mbox{$\xi (\s)\in C_{(2)}(K(X),\Z)$} given by (\ref{xis}). It turns out that
small deformations have a very simple interpretation in terms of
chains. If $\s_1$ and $\s_2$ differ
by a minimal deformation of the type illustrated in Fig. 1(d),
it follows from the definitions that  $\xi (\s_1)-\xi (\s_2)$ is
equal to zero. In general we have
\be
\s_1\sim \s_2~~\mbox{ implies } ~~\xi (\s_2)= \xi (\s_2)+\partial B,
\label{5.14}
\ee
for some 3-chain $B$.

Equation (\ref{5.14}) is the clue to the invariance of $I[\s]$ under
continuous deformations. In order to proceed it will be useful to introduce
two kinds of products involving chains. The first one is a scalar product
$\langle \cdot,\cdot \rangle $ on $C(K(X),\Z)$. As usual it is enough to give
the product for the base elements. We define
\be
\langle \a_{(m)}^i,\a_{(n)}^j\rangle = \delta _{mn}\delta _{ij}.\label{inter}
\ee
The second one is a cross product.
Let $\a_{(m)}^i$ and $\a_{(n)}^j$ be base elements such that \mbox{$(m+n)=4$}.
The cross product $\a_{(m)}^i\times \a_{(n)}^j$ will be
given by (\ref{x}) if they belong to the same $4$-cell, and will be zero
otherwise. For arbitrary chains $\xi _{(m)}^1$ and $\xi _{(n)}^2$ the
product is extended by linearity.

Let us regard the cell decomposition $K(X)$ as a vector on $C_{(4)}(K(X),\Z)$.
We can assume that all 4-cells $\a_{(4)}^i$ have a coherent orientation and
therefore
\be
K(X)=\sum _i \a_{(4)}^i, \label{eq1}
\ee
where the sum is over all 4-cells. Let $\s$ be a transversal configuration with
$n$ plaquettes.
In this case, the expression (\ref{xis}) reduces to
\be
\xi (\s)= s_1\a_{(2)}^1+s_2\a_{(2)}^2+...+s_n\a_{(2)}^n.\label{eq2}
\ee
Consider the product $\xi (\s)\times \xi (\s)$. Because of
the way the cross product was defined, most of the $n^2$ terms
in the expansion of $\xi (\s)\times \xi (\s)$ will be zero.
There will be contributions only from plaquettes that share exactly one
vertex, or in other words, from plaquettes that contain the
intersection points $v$.
It is not difficult to see that there will be 32 non
vanishing terms per each intersection point $v$. From (\ref{x}),
(\ref{7.2}) and (\ref{ss}) one can show that each term is equal a 4-cell
multiplied by the intersection number $I[\s_v,\s_v']$.
Combining (\ref{eq1}) and (\ref{eq2}) with the previous observation, one can
see that the oriented self-intersection number $I[\s]$ can
be expressed as
\be
I[\s]=\frac{1}{32}\langle K(X),\xi (\s)\times \xi (\s)\rangle .\label{7.3}
\ee

The topological invariance of (\ref{7.3}) is a consequence of the identity
\be
\langle K(X),\xi _{(2)} \times \partial \xi _{(3)}\rangle=
\langle K(X),\partial \xi _{(2)} \times \xi _{(3)}\rangle , \label{7.4}
\ee
where $\xi _{(2)}$ and $\xi _{(3)}$ are arbitrary 2-chains and 3-chains.
Let us assume (\ref{7.4}) for the moment. Given $\s \sim \s'$, it
follows from (\ref{5.14}), (\ref{7.3}) that
\be
I[\s']=\frac{1}{32} \langle K(X), \xi (\s) \times \xi (\s) \rangle +
\frac{1}{16} \langle K(X), \xi (\s) \times \partial B\rangle +
\frac{1}{32} \langle K(X),  \partial B\times \partial B\rangle. \label{top0}
\ee
Using the identities (\ref{7.4}) and $\partial ^2=0$ we have
\be
I[\s']-I[\s]=\frac{1}{16} \langle K(X), \partial \xi (\s) \times B\rangle
\label{top}
\ee
If $\s$ has no boundary, then $\partial \xi (\s)=0$ and $I[\s]=I[\s']$. When
the surface $\s$ has a boundary, (\ref{7.3}) is still well defined, but is no
longer invariant under arbitrary deformations.
One has to be restricted to the class of  deformations such that
$\langle K(X),\partial \xi (\s) \times B\rangle=0$. For example,
if the fluctuations on $\s$ occur far from its boundary, i.e.,
$B=0$ at the boundary of $\s$, the r.h.s. of (\ref{top}) obviously vanishes.

The extension of $I[\s]$ to non-transversal configurations is now obvious.
Given $\s$, one computes $\xi (\s)$ by formula (\ref{xis}) and uses
(\ref{7.3}) to compute $I[\s]$. This is a well-defined procedure, since the
r.h.s. of (\ref{7.3})
makes sense for an arbitrary vector $\xi (\s)$ in $C_{(2)}(K(X),\Z)$.

We would like to indicate how identity (\ref{7.4}) can be proven. It is
enough to verify it for the base elements in $C_{(2)}(K(X),\Z)$ and
$C_{(3)}(K(X),\Z)$. In the notation of  \mbox{Section \ref{se:4.2}}, let
\begin{eqnarray}
\xi _{(2)}&=&\left[ (l^a,l^b,v^c,v^d)\right] \nonumber \\
\xi _{(3)}&=&\left[ (\tilde l^a,\tilde l^i,\tilde v^j,\tilde v^k)\right].
\end{eqnarray}
Notice that $\xi _{(2)}$ and $\xi _{(3)}$ have to have link components on one
common direction given by the repeated index $a$. Let us first compute
$\langle K,\partial \xi _{(2)}\times \xi _{(3)}\rangle$. The only terms in
$\partial \xi _{(2)}$ that contribute are
\[
\e ^{ab}\left[ ([p^a],l^b,v^c,v^d)\right]-
   \e^{ab}\left[ ([p^a+1],l^b,v^c,v^d)\right].
\]
After some algebra we have
\be
\langle K(X),\partial \xi _{(2)}\times \xi _{(3)}\rangle =
\e^{ab}\e^{[ajk]b}\left(\d _{[p^a+1]\subset \tilde l^a}-
                 \d _{[p^a]\subset \tilde l^a} \right) ,\label{A}
\ee
where $\d _{[p^a]\subset \tilde l^a}$ is zero unless $[p^a]\subset \tilde l^a$.
Analogously,
\be
\langle K(X),\xi _{(2)}\times \partial \xi _{(3)}\rangle =
\e^{a[jk]}\e^{[ab][jk]}\left( \d _{[\tilde p^a]\subset l^a}-
			\d _{[\tilde p^a+1]\subset l^a}\right) .\label{B}
\ee
The fact that $K(X)$ has no boundary has been used to derive the
last two equations. A little thought shows that
$\e^{ab}\e^{[ajk]b}=\e^{a[jk]}\e^{[ab][jk]}$.
If $l^a$ and $\tilde l^a$ do not share any vertex or if
$l^a=\tilde l^a$, (\ref{A}) and (\ref{B}) are both zero. For the cases where
they are adjacent, the r.h.s. of (\ref{A}) and (\ref{B}) give the same result.

\sxn{Final Remarks}

A discrete model of random surfaces with topological term was
introduced. The model is described by the Green functions
$Z_{n,m}(\g_1,...,\g_n;\l ,\theta )$ defined by (\ref{f.1}). We show that
the topological term $I[\s]$ is well defined and can be
computed explicitly by formula (\ref{7.3}) for the cases where the target
space $X$ is $\R^4$ or the 4-torus $\T^4$.

In principle, one can study the behavior of
$Z_{n,m}(\g_1,...,\g_n;\l \theta )$  for a fixed number $m$ of handles.
Let us examine the case $n=m=0$ and $X=\R^4$. The partition function
$Z_{0,0}$ is a sum over
surfaces with the topology of $S^2$. Since $\s$ has no
boundary, the corresponding chain $\xi (\s)$ is actually a cycle, i.e.
\be
\partial \xi(\s)=0 \label{f.2}
\ee
But $K(X=\R^4)$ is homologicaly trivial, and all closed chains are also
exact \cite{HW}. Therefore
\be
\xi(\s)=\partial \omega \label{f.3}
\ee
for some $\omega \in C_{(1)}(K(X),\Z)$.
{}From (\ref{7.3}), (\ref{7.4}) and (\ref{f.3}) one sees immediately that
\be
I[\s]=0.\label{f.4}
\ee
Even though $\s$ can self-intersect at many points, the intersection numbers
add up to zero\fn{In particular, for transversal configurations the total
number of intersection points has to be even.}. In the computation of
$Z_{0,0}$, it does not matter if  the  $\theta $-angle is zero or not. In
other words
\be
Z_{0,0}(\l,\theta )=Z_{0,0}(\l,0)
\ee

Contrary to $Z_{0,0}$, the "$n$ point" Green functions $Z_{n,0}$ depend on
$\theta $. For example, let
us examine $Z_{1,0}$. The sum is now performed over the set $\G_d(\g)$ of
surfaces $\s$ with boundary $\g$ and no handles, in other words, surfaces with
the topology of a disk.
In contrast with (\ref{f.4}), one can easily show that there are surfaces
in $\G_d(\g)$ that have self-intersection numbers different from zero.

Consider the following construction.  Let $\s_0$ be a transversal
surface with the topology of $S^2$. Suppose  that $\s_0$ self-intersects at
$2k$ points $v_i\in K(X)$. For each $v_i$ there is a corresponding pair
of points $p_i$ and $p_i'$ in the abstract cell complex $\s_0$.
Let us associate a ``charge'' $\pm \frac12$ to each
point $p_i$ and $p_i'$ according if the intersection number at $v_i$ is
$\pm 1$. From (\ref{f.4}) it follows that the total ``charge'' is zero.
Consider now a loop $\g$ dividing $\s_0$ into disks $\s_1$ and $\s_2$ with
$\partial \s_1=\partial (\s_2)=\g$. Some pairs of points $p_i, p_i'$ will be
completely contained in $\s_1$ and some others will
have one point in $\s_1$
and the other point in $\s_2$. (We assume that $\g$ does not touch any
intersection.) Let us call $q_i$ the ``charge'' in
$\s_i$ ($i=1,2$) due to the pairs that are not divided by
$\g$, and $q_{12}$ the  remaining ``charge''. Then
\be
q_1+q_{12}+q_2=0.\label{c}
\ee
The intersection number $I[\s_i]$ is obviously equal to $q_i$. Then, the
contribution of $\s_1$ and $\s_2$ to $Z_1(\g;\theta )$ is given by
\be
e^{iq_1\theta }e^{-\l A[\s_1]}+ e^{iq_2\theta }e^{-\l A[\s_2]} \label{f.5}
\ee
In particular, if $\theta =\p$, $q_1$ is even and $q_2$ is odd, the
contribution (\ref{f.5}) reduces to
\be
e^{-\l A[\s_1]}-e^{-\l A[\s_2]}.
\ee
{}From (\ref{c}), one can see that $q_1$ and $q_2$ are integers such that
$-k\leq q_1+q_2\leq k$. Therefore, $Z_1(\g;\l,\theta )$ does depend on
$\theta $.

Nothing much is known about the critical behavior of the model in the entire
parameter space $(\l ,\theta )$, except for $\theta =0$.
Unfortunately, for $\theta =0$ the continuum limit is trivial.
The sickness of the model at $\theta =0$ is a
consequence of the fact that the bare string tension has no
zeros \cite{DFJ2}. However, due to (\ref{f.5}), the bare string tension
\[
\t(\l,\theta )=\lim _{\g\rt \infty}\frac{1}{LM}\log
{Z_{1,0}(\g_{LM};\l,\theta )}.
\]
can have a radically different behavior  for $\theta \neq 0$. It is
conceivable that, for $\theta =\p$, there are critical points where
$\t(\l ,\theta )$ does go to zero. The speculation of such a nontrivial
continuum limit deserves further investigation.

\vspace{2cm}
\noindent
{\Large \bf Acknowledgments }

I would like to thank M. Bowick for bringing to my attention the
problem of the self-intersection number and for many useful discussions.
I also would like to thank A.P. Balachandran, L. Chandar, E. Ercolessi,
G. Harris and S. Vaidya for their comments and suggestions.
This work was supported by the Department of Energy under contract number
DE-FG-02-84ER40173.

\end{document}